\documentclass[aps,floatfix,prb,twocolumn,a4paper,10pt,citeautoscript,showpacs,
superscriptaddress]{revtex4-1}

\usepackage{bm}
\usepackage{amsmath}		
\usepackage{amssymb}		

\usepackage{graphicx}	
\usepackage{graphics}	
\DeclareGraphicsExtensions{.pdf}	

\usepackage[colorlinks=true, pdfstartview=FitV,linkcolor=blue, citecolor=blue, urlcolor=blue]{hyperref}

\newcommand{\vnabla}{{\mbox{\boldmath$\nabla$}}}

\newcommand{\vR}{{\mbox{\boldmath$R$}}}

\newcommand{\vk}{{\mbox{\boldmath$k$}}}
\newcommand{\vv}{{\mbox{\boldmath$v$}}}

\newcommand{\vsk}{{\mbox{\small\boldmath$k$}}}

\newcommand{\hvk}{\hat{\mbox{\boldmath$k$}}}

\newcommand{\vsig}{\mbox{\boldmath$\sigma$}}

\makeatletter
\newcommand*{\Equation}{\@ifstar\sEquation\oEquation}
\newcommand{\sEquation}[1]{\begin{equation*}#1\end{equation*}}
\newcommand{\oEquation}[2]{  \begin{equation}\label{#1}#2\end{equation} }
\makeatother
\newcommand{\Align}[2]{\begin{align}\label{#1}#2\end{align}}
\newcommand{\Ref}[1]{\ref{#1}}
\newcommand{\Eqref}[1]{\eqref{#1}}
\newcommand{\Figref}[1]{Fig.~\ref{#1}}
\newcommand{\Hc}[1]{\mathrm{H}_{c#1}} 
\newcommand{\Q}{\mathcal{Q}}
\newcommand{\groupU}[1]{\mathrm{U}(#1)} 
\newcommand{\groupCP}[1]{{\mathbb{C}{P}}^{#1}} 
\newcommand{\CPone}{$\groupCP{1}\ $}
\newcommand{\Uone}{$\groupU{1}\ $}
\newcommand{\groupUU}{$\groupU{1}\times\groupU{1}\ $} 

\renewcommand\Im{\mathrm{Im}}

\newcommand{\Grad}{\bs \nabla}

\newcommand{\Curl}{\bs \nabla\times}

\newcommand{\bs}{\boldsymbol}
\newcommand{\D}{{\bs D}}

\newcommand{\A}{{\bs A}}
\newcommand{\B}{{\bs B}}

\newcommand{\J}{{\bs J}}

\newcommand{\psia}{\psi_{a}}

\begin{document}

\title{Microscopic prediction of skyrmion lattice state in clean interface superconductors}
\author{Daniel~F.~Agterberg}
\affiliation{Department of Physics, University of Wisconsin-Milwaukee, Milwaukee, WI 53211}
\author{Egor~Babaev}
\affiliation{Department of Theoretical Physics, Royal Institute of Technology, Stockholm, SE-10691 Sweden}
\affiliation{Department of Physics, University of Massachusetts Amherst, MA 01003 USA }
\author{Julien~Garaud}
\affiliation{Department of Physics, University of Massachusetts Amherst, MA 01003 USA }
\affiliation{Department of Theoretical Physics, Royal Institute of Technology, Stockholm, SE-10691 Sweden}

\begin{abstract}
When an in-plane field is applied to a clean interface superconductor, a Fulde-Ferrell-Larkin-Ovchinnikov (FFLO)-like phase is stabilized. This phase has a \groupUU symmetry and, in principle, this symmetry allows for flux carrying topological excitations different from Abrikosov vortices (which are the simplest defects associated with $S^1 \to S^1$ maps). However, in practice, largely due to electromagnetic and other intercomponent interactions,  such topological excitations are very rare in superconducting systems. Here we demonstrate that a realistic microscopic theory for interface superconductors, such as SrTiO$_3$/LaAlO$_3$, predicts an unconventional magnetic response where the flux-carrying objects are skyrmions, characterized by homotopy invariants of $S^2 \to S^2$ maps. Additionally, we show that this microscopic theory predicts that stable fractional vortices form near the boundary of these superconductors. It also predicts the appearance of type-1.5 superconductivity for some range of parameters. Central to these results is the assumption that the Rashba spin orbit coupling is much larger than the superconducting gap.
\end{abstract}
\maketitle

\section{Introduction}
Since the original discovery of superconductivity at the interface of SrTiO$_3$ and LaAlO$_3$ \cite{oht04,rey07}, the field of two-dimensional (2D) interface superconductors has grown tremendously.  Some notable examples include electric field induced superconductivity in SrTiO$_3$ \cite{uen08}, KTaO$_3$ \cite{uen11}, and  MoS$_2$ \cite{tan12,ye12}.  This growing set of materials present an ideal opportunity to examine new physics associated with the superconducting state. One predication for clean interface superconductors with a large Rashba spin-orbit coupling (large with respect to the superconducting gap) is that a Fulde-Ferrell-Larkin-Ovchinnikov (FFLO)-like \cite{ff,lo} phase appears with in-plane fields \cite{bar02,agt03,dim07,agt07}. This phase is more robust than the usual FFLO phase \cite{bar02,agt03,dim07,sam08,mic12}. In this phase, the superconducting order breaks translational symmetry in addition to gauge symmetry, and consequently, the order parameter has a \groupUU symmetry. In principle such a symmetry allows non-trivial topological defects, such as fractional vortices, which give rise to interesting consequences \cite{npb,agt08,rad09,rad12,gop13,agt08-2,ber09,ber09-2,agt11,bab02,sil11,Garaud.Sellin.ea:14}. However, it is unclear that these non-trivial defects are energetically stable. For example, fractional vortices  are energetically more expensive than integer-flux vortices and thus are typically excluded from the magnetic response under normal conditions \cite{bab02}. A different  class of  topological defects is also theoretically possible when the minimum energy excitation is neither a fractional nor a usual integer-flux vortex,  but a skyrmion. Skyrmions can be visualized as states in which pairs of fractional vortices have a preferred separation.  As discussed in detail below such configurations, carrying integer flux quanta are topologically distinct from Abrikosov vortices since they are characterized by homotopy invariants of the maps $S^2\to S^2$. Although the existence of such states in superconductors can be justified by symmetry-based arguments, they have not been experimentally observed nor do they have a microscopic basis. Here we give such a microscopic justification. In particular, we show that when a $c$-axis field is applied in the FFLO-like phase, such flux-carrying skyrmion defects are ubiquitous.

We consider a weak-coupling theory for a clean superconductor with isotropic pairing interactions ($s$-wave pairing), an in-plane Zeeman field, and a strong Rashba spin-orbit coupling (with respect to the Zeeman field and the superconducting gap). We consider the chemical potential to be well above the Dirac point, away from the limit at which Majorana modes are predicted. For a wide range of in-plane fields and temperatures, we show there exists phases akin to the FFLO phase \cite{agt07,dim07}.  Specifically, we find two FFLO-like phases: a single-$Q$ phase (also known as the FF phase) in which the superconducting gap function takes the form $\Delta({\vR})=\Delta_qe^{iqx}$ and the multiple-$Q$ phase which can be qualitatively be described by the gap function $\Delta({\vR})=\Delta_qe^{iqx}+\Delta_{-q}e^{-iqx}$ with both $\Delta_q$ and $\Delta_{-q}$ non-zero and $\Delta_q\ne \Delta_{-q}$. We show that the phase boundaries for these FFLO-like phases are well described by a Ginzburg Landau (GL) theory derived from this weak coupling theory. Using this GL theory, we find our central result: fractional vortices and skyrmions (split-core vortices) are prevalent in the multiple-$Q$ phase when a magnetic field is applied perpendicular to the plane. We further show that the multiple-$Q$ phase must also  exhibit type-1.5 superconductivity in a region of the phase diagram. In this regime there are coherence lengths that are larger and smaller than the magnetic field penetration length.  This can result in a  coalescence and formation of vortex aggregates surrounded by vortexless regions.

This paper is arranged as follows.
In Sec.~\ref{Sec:General} we discuss the phenomenology of the \groupUU
Ginzburg-Landau model and its topological excitations. Sec.~\ref{Sec:Micro}
is devoted to the microscopic theory that gives rise to the FFLO-like phases.
In Sec.~\ref{Sec:Inplane}, we show that for some range of the temperature
and in-plane field, this microscopic theory can be well approximated by a \groupUU
Ginzburg-Landau theory whose phenomenological properties where previously
introduced in Sec.~\ref{Sec:General}. Sec.~\ref{Sec:C-axis} is devoted to
the numerical study of topological defects that appear in response to fields
applied along the $c$-axis, in the multiple-$Q$ phase.

\section{Fractional vortices and skyrmions in \texorpdfstring{\groupUU}{U(1)xU(1)} theories.}
\label{Sec:General}
The GL theory for the order parameter, derived below, takes the form
\begin{align}
F&= \int d^2x \Bigg\{ \frac{1}{2}B^2-H_zB
+\beta_m|\Delta_q|^2|\Delta_{-q}|^2	\nonumber	\\
+&\sum_{i=x,y}\kappa_{1i}|D_i\Delta_q|^2
+\alpha_1|\Delta_q|^2+\frac{\beta_1}{2}|\Delta_q|^4
\nonumber \\
+&\sum_{i=x,y}\kappa_{2i}|D_i\Delta_{-q}|^2
+\alpha_2|\Delta_{-q}|^2+\frac{\beta_2}{2}|\Delta_{-q}|^4
\Bigg\} \,,\label{GL-free}
\end{align}
where $D_j=\partial_j+2ie A_j$, $B=\partial_x A_y-\partial_y A_x$, and $H_z$
is the applied field normal to the interface.
$\Delta_{\pm q}=|\Delta_{\pm q}|\mathrm{e}^{i\varphi_\pm}$ are complex fields
representing the superconducting condensates. The free energy has a \groupUU
invariance. In the context of the microscopic theory discussed below, one 
\Uone symmetry is the usual gauge invariance while the second \Uone symmetry 
stems from translational invariance.
Depending on the parameters $\alpha_1$, $\beta_1$, $\alpha_2$, $\beta_2$
and $\beta_m$ of the interacting potential, there can be three homogeneous
ground states. Two of these we name the single-$Q$ phase, in which either
$\Delta_q=0$ or $\Delta_{-q}=0$, and the third is the multiple-$Q$ phase
(occurring when $\beta_1\beta_2>\beta_m^2$ and $\alpha_1,\alpha_2<0$)
for which both $\Delta_q\ne 0$ and $\Delta_{-q}\ne 0$.
There are two different coherence lengths $\xi_\pm$. These can be uniquely
associated with the condensates $\Delta_{\pm q}$ in the single-$Q$ phase.
In the multiple-Q phase, these two coherence lengths describe linear 
combinations of  $\Delta_{\pm q}$. $\xi_{+}<\xi_-$ for the entire range 
of parameters and $\xi_-$ diverges at the single-$Q$ to multiple-$Q$ 
transition (see details in Appendix \ref{Appendix}).

The parameters $\alpha$'s, $\beta$'s and $\kappa$'s of the model are 
microscopically determined and $e$, which is used to parametrize the 
penetration depth of the magnetic field, is the only free parameter
of the GL free energy Eq.~\Eqref{GL-free}.
Because of the \groupUU symmetry, both condensates are independently
conserved. This implies that, in general, there will be two
second critical fields $\Hc{2}^{(-q)}<\Hc{2}^{(q)}$ associated with
the destruction of the corresponding condensates.

The elementary topological excitations here are fractional vortices.
That is, field configurations with independent $2\pi$ windings in either
$\Delta_{\pm q}$ (e.g. $\varphi_+$ has $\oint\nabla\varphi_+=2\pi$
winding while $\oint\nabla\varphi_-=0$). Configurations with winding
$n$ in $\Delta_{q}$ and $m$ in $\Delta_{-q}$, denoted $(n,m)$, carry
a flux that is not necessarily an integer multiple of the flux quantum
$\Phi_0=2\pi/e$ \cite{bab02},
\begin{equation}
\Phi_{n,m}=\frac{n\kappa_{1x}|\Delta_q|^2+m\kappa_{2x}|\Delta_{-q}|^2}
{\kappa_{1x}|\Delta_{q}|^2+\kappa_{2x}|\Delta_{-q}|^2}\Phi_0 \,.
\label{flux}
\end{equation}
In the single-$Q$ phase, where only one component condenses, the flux
carried is always an integer multiple of $\Phi_0$. In the multiple-$Q$
phase, each component $\Delta_{q}$ and $\Delta_{-q}$ respectively carries 
$\Phi_{1,0}$ and $\Phi_{0,1}$, fractions of the flux quantum. When both 
condensates have the same winding $m=n$, both these fractions add up to 
an integer multiple of $\Phi_0$. The corresponding configurations are $n$ 
``composite'' vortices, each carrying one flux quantum. These have finite 
energy per unit length (independent of system size) due to screened currents.
A detailed derivation of the flux quantization and fractional flux
carried by the different condensates is given in Appendix~\ref{Appendix}.
When there are fractional vortices, that is when $m\ne n$, there are unscreened 
counter-currents in both components. This leads to logarithmically divergent 
energy per unit length, making their creation in the bulk unlikely. Nevertheless, 
fractional vortices can be thermodynamically stable near boundaries \cite{sil11}.

Typically, a \groupUU superconductor in an external field forms composite
vortices with $(1,1)$ winding due to the logarithmic attraction of fractional
$(1,0)$ and $(0,1)$ vortices.
Winding in the relative phase $\varphi_r=\varphi_+-\varphi_-$ signals 
fractional vortices. That is, local $2\pi$ winding in $\varphi_r$ signals
non overlapping cores of fractional vortices in both condensates. This can
be due to either fractional vortices or to split-core vortices, skyrmions,
as discussed below.
Although two-component superconductors are common, the associated skyrmions
are usually unstable and coalesce into Abrikosov vortices. We show below that 
in the multiple-$Q$ phase of the microscopic theory, skyrmions commonly appear
in an external field.

\section{Microscopic formulation.}
\label{Sec:Micro}
We consider the following Hamiltonian:
\begin{align}
{\cal H}&= \sum_{{\bf k},\sigma}a^{\dagger}_{{\bf k}\sigma}\xi_{{\bf k}}a_{{\bf k}\sigma} + \sum_{{\bf k} \sigma \sigma'} a^{\dagger}_{{\bf k} \sigma}[\alpha{\bf g}_{\bf k}+\mu_B{\bf H}]\cdot \vsig_{\sigma \sigma'}a_{{\bf k}\sigma'} \nonumber\\
&+\frac{1}{2}V\sum_{{\bf k},{\bf k}',{\bf q}}a^{\dagger}_{{\bf k}+{\bf q}\uparrow}a^{\dagger}_{-{\bf k}+{\bf q}\downarrow}a_{-{\bf k}'+{\bf q}\downarrow}a_{{\bf k}'+{\bf q}\uparrow}
\label{eq1}
\end{align}
where $a_{{\bf k} \sigma}$ are the annihilation operators with momentum ${\bf k}$ and with pseudospin $\sigma$, $\xi_{\bf k}=\epsilon_{\bf k}-\mu$, ${\bf g}_{\bf k}=(k_y,-k_x)/k_F$ is the Rashba spin-orbit coupling, ${\bf H}$ is an in-plane Zeeman field, $\epsilon_{\bf k}=k^2/(2m)$, and $V$ is the $s$-wave pairing interaction. We assume a large Rashba spin-orbit coupling, so that $T_c,|\mu_B {\bf H}|\ll|\alpha|$.  We ignore terms of the order $(\mu_B |{\bf H}|/\alpha)^2$, this limit considerably simplifies the theory. The eigenstates of the single-particle Hamiltonian are given by the helicity basis. Specifically, helicity annihilation operators $a_{{\bf k}\pm}$ are given by $a_{{\vsk}\alpha}=\sum_s U_{\alpha,s}(k)c_{{\vsk}s}$, with $U(k)=\frac{1}{\sqrt{2}}[1-i(\cos\tilde{\phi} \sigma_y-\sin\tilde{\phi} \sigma_x)]$, $\alpha{\bf g}_{\bf k}+\mu_B{\bf H}=|\alpha{\bf g}_{\bf k}+\mu_B{\bf H}|(\cos \tilde{\phi},\sin\tilde{\phi})$, and $a_{{\bf k}\pm}$
annihilates particles in the two spin-dependent bands with energies $E_{\vk,\pm}=\epsilon(\vsk)\pm |\alpha{\bf g}_\vsk+\mu_B{\bf H}|$. The description of superconductivity in the helicity basis is akin to a two-band theory for which we can follow standard methods to find the corresponding Eilenberger equations that describe the weak-coupling limit \cite{eil68,lar68,ser83}.
 This limit assumes $k_BT_c\ll\omega_c\ll\epsilon_F$ and $1/k_F\xi_0\ll1$ where $T_c$ is the transition temperature, $\omega_c$ is the cut-off frequency, $\epsilon_F=\hbar^2 k_F^2/2m$ is the Fermi energy, and $\xi_0$ is zero-temperature superconducting coherence length.

To find the Eilenberger equations, we define the usual Green's functions in Nambu space for each helicity band ${\bm \Psi}_{\pm}^\dagger(\bm x)=[\psi^\dagger_{\pm}(\bm x),\psi_{\pm}(\bm x)]$ and define the imaginary time Green's function as
\begin{equation}
\hat{G}_{\pm}({\bm x}_1,{\bm x}_2;\tau_1-\tau_2)=-\langle T_\tau {\bm \Psi}_{\pm}({\bm x}_1,\tau_1) {\bm \Psi}_{\pm}^\dagger({\bm x}_2,\tau_2)\rangle, \label{greens-func}
\end{equation}
here the operator $T_\tau$ arranges the field operators in ascending order of the imaginary time $0<\tau<1/T$ and $\bm{\Psi}({\bm x},\tau)=e^{\tau \cal{H}}\bm{\Psi}(\bm x)e^{-\tau\cal{H}}$. Introducing the center-of-mass coordinate, ${\bm R}=({\bm x}_1+{\bm x}_2)/2$, the relative coordinate, ${\bm r}={\bm x}_1-{\bm x}_2$, and performing the Fourier transform in the latter variable yields
\begin{equation}
\hat{G}_{\pm}(\vk,\vR;\omega_n)=\int d{\bm r}\int_0^{1/T} d{\tau} \hat{G}_{\pm}\left({\bm x}_1,{\bm x}_2;\tau\right) e^{-i({\bm k}\cdot{\bm
r}-\omega_n\tau)},
\end{equation}
where $\omega_n=(2n+1)\pi T$ is the fermionic Matsubara frequency.
We define the quasi-classical Greens functions
\begin{equation}
\hat{g}_{\pm}(\hvk,\vR,\omega_n)= \left(\begin{array}{cc}
g_{\pm} & f_{\pm} \\
f^\dagger_{\pm} & -g_{\pm}
\end{array}\right)
\equiv\frac{i}{\pi}\int d\xi\hat{\tau_3}\hat{G}_{\pm}(\vk,\vR,\omega_n),
\end{equation}
where $d\xi$ integrates out the variable  perpendicular to the Fermi surface, $\hvk$ is vector on the Fermi surface, and $\tau_3$ is the $z$-component of the Pauli matrices acting on the particle-hole space.

The standard quasi-classical approach \cite{eil68,lar68,ser83} results in the following Eilenberger equations for this system (here we have assumed a small Zeeman field, that is we have kept terms up to order $\mu_B |{\bf H}|/\alpha$):
\begin{equation}
[\omega_n\pm i \mu_B\hat{z}\cdot \hat{{\bf k}}\times {\bf H}+\frac{1}{2} \vv_{\hvk}\cdot(\vnabla-2ie{\bf A})]f_{\pm}=\Delta_{\pm}(\hvk,\vR)g_{\pm} \label{eilen1}
\end{equation}
\begin{equation}
[\omega_n\pm i\mu_B \hat{z}\cdot \hat{{\bf k}}\times {\bf H} -\frac{1}{2} \vv_{\hvk}\cdot(\vnabla+2ie{\bf A})]f_{\pm}^{\dagger}=\Delta^*_{\pm}(\hvk,\vR)g_{\pm}
\label{eilen2}
\end{equation}
where $f_{\pm}^{\dagger}f_{\pm}+g_{\pm}^2=1$ and $\vv_{\hvk}$ is the Fermi velocity (within the approximations used here, $\vv_{\hvk}$ is the same for both bands and is independent of ${\bf H}$). The gap equation is
\begin{equation}
\Delta_i(\hvk,\vR)=\pi T\sum_{n,j}N_j\left\langle \tilde{V}_{ij}(\hvk,\hvk')f_j(\hvk',\vR,\omega_n)\right\rangle_{\hvk'}
\end{equation}
where $N_j$ is the density of states on band $j$, $\left\langle f\right\rangle_{\hvk'}$ means average $f$ over $\hvk'$, and the effective two-band pairing interaction $\tilde{V}$  is (with a finite Zeeman field, this remains correct up to order $(\mu_B |{\bf H}|/\alpha)^2$)
\begin{equation}
\tilde{V}=\frac{1}{2}V\left(
                          \begin{array}{cc}
                            e^{i(\phi-\phi')} & e^{i(\phi+\phi')} \\
                            e^{-i(\phi+\phi')} & e^{-i(\phi-\phi')}\\
                          \end{array}
                        \right)
                        \label{pair-hel}
\end{equation} where $e^{i\phi}=(k_x+ik_y)/|{\bf k}|$.
Redefining the gap functions $\Delta_\pm(\vk,\vR)=\pm e^{\pm i\phi}\tilde{\Delta}_\pm(\vk,\vR)$ and the propagators $f_\pm(\vk,\vR,i\omega_n)=\pm e^{\pm i\phi}\tilde{f}_\pm(\vk,\vR,i\omega_n)$ yields simplified
two-band Eilenberger equations. The gap functions can then be written as
$\tilde{\Delta}_{\alpha}(\hvk,\vR)=\Delta_{\alpha}(\vR)$
and the gap equation  with the form of $V$ in Eq.~\Eqref{pair-hel} implies that
$\Delta_{+}(\vR)-\Delta_{-}(\vR)=0$.
In the following, we set $\Delta(\vR)=\Delta_+(\vR)=\Delta_-(\vR)$ and
$f_\alpha(\hvk,\vR,\omega_n)=\tilde{f}_\alpha(\hvk,\vR,\omega_n)$.

The Eilenberger equations can be derived from a Gibbs free-energy
functional \cite{eil68}. Once the Eilenberger equations are solved 
for a given functional form of $\Delta(\vR)$ and a known magnetic field, 
this free energy functional becomes
\begin{align}
&\Omega_{\rm SN}= \int d{\bm R} \Bigg[V^{-1} \left|\Delta(\vR)\right|^2 \nonumber  \\
& -\pi T \sum_{n\ge 0,j}N_j \left\langle I_j(\hvk,\vR,\omega_n)+I^*_j(\hvk,\vR,\omega_n)\right\rangle\Bigg],
\label{free-energy}
\end{align}
where
\begin{equation}
I_j(\hvk,\vR,\omega_n)= \frac{{\Delta}^*(\bm R)f_j(\hvk,\vR,\omega_n) +f^\dagger_j(\hvk,\vR,\omega_n){\Delta}({\bm
R})}{1+g_j(\hvk,\vR,\omega_n)}. \label{func-i}
\end{equation}

Prior to proceeding, it is useful to define $\delta N$ which plays a central role in the theory. $\delta N$ quantifies the difference in the density of states on the two Fermi surfaces associated with different helicity, it is defined as
\begin{equation}
\delta N= \frac{N_+-N_-}{N_++N_-}=\frac{\alpha}{2\epsilon_F}.
\end{equation}

\begin{figure*}
\vspace{-0.75cm}
\hbox to \linewidth{ \hss
\includegraphics[width=0.95\linewidth]{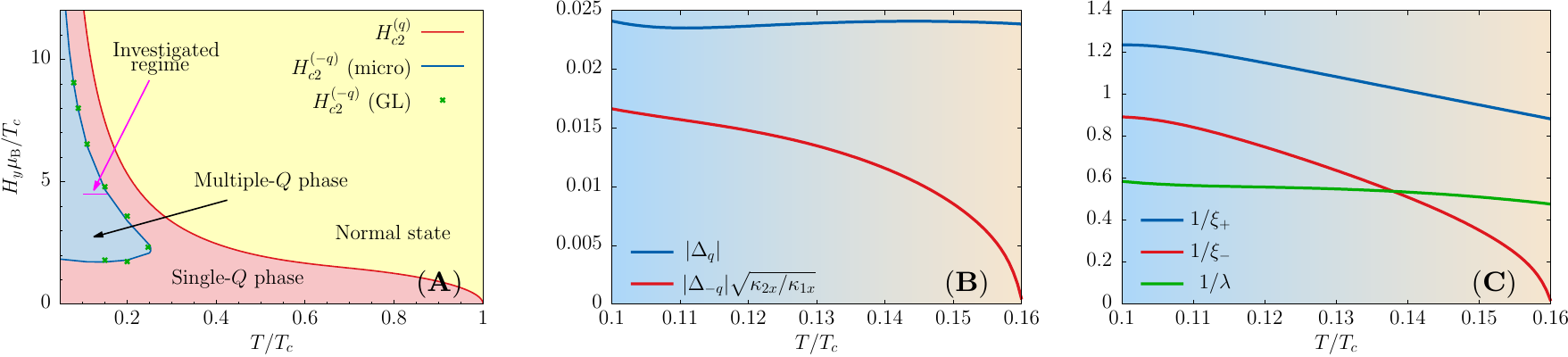}
\hss}
\vspace{-0.3cm}
\caption{
Panel (A) shows the temperature-magnetic field phase diagram for an $s$-wave superconductor with $\delta N=0.05$ (for in-plane magnetic fields). The  line $H_{c2}^{(q)}$ denotes the transition from the normal phase to single-$Q$ superconductor. The line $H_{c2}^{(-q)}$(micro) denotes the multiple-$Q$ to single-$Q$ phase transition found using the microscopic theory described in the text. The green crosses denote this same transition within the Ginzburg-Landau approximation of the microscopic theory. The line labeled ``Investigated regime'' is where we examined the role of an out of plane magnetic field ($H_y\mu_B/T_c=4.5$ along this line).
Panels (B) and (C), respectively, display the ground-state densities and the
length scales, in the investigated regime. That is, in multiple-$Q$ phase,
where both components $\Delta_{q}$ and $\Delta_{-q}$ have non zero ground
state densities. 
Note that the (hybridized) coherence length  $\xi_-$ diverges when the 
$\Delta_{-q}$ vanishes, while other length scales remain finite. Thus, in the 
vicinity of the temperature $T/T_c=0.16$, the penetration depth $\lambda$ (here 
$4e\sqrt{\kappa_{1x}\kappa_{1y}}=0.05$), can be an intermediate length scale 
(see discussion in Sec.~\ref{type15}). 
}
\label{Fig:phase}
\end{figure*}

\section{Phase diagram for in-plane fields.}
\label{Sec:Inplane}
We consider a field along the $\hat{y}$ direction. Phenomenological arguments imply that $\Delta({\bf x})=\Delta_q e^{iqx}$ close to normal to superconducting transition \cite{agt03}. We have examined the stability of this single-$Q$ solution within the microscopic theory discussed above. In particular, we have minimized the microscopic free energy in Eq.~\Eqref{free-energy} with respect to $q$ and $\Delta_q$. Once the optimal single-$Q$ solution has been found we then set ${\Delta}({\bf x})=\Delta_qe^{iqx}+\Delta_{-q}e^{-iqx}+\Delta_{3q}e^{i3qx}$ and expand the free energy to quadratic order in $\Delta_{-q}$ and $\Delta_{3q}$. If this additional contribution lowers the free energy, then the multiple-$Q$ phase is stable. Our approach agrees with earlier calculations done for $\delta N=0$ \cite{dim07}. For $\delta N<0.25$, we find that the phase diagram generically contains both single-$Q$ and multiple-$Q$ phases and resembles that shown in \Figref{Fig:phase}. For $\delta N\ge 0.25$, we find only the single-$Q$ phase is stable.

We have also derived the GL free energy of Eq.~\Eqref{GL-free} from the microscopic theory ignoring $\Delta_{3q}$ (which numerically is found to be small). As shown in Fig.~\ref{Fig:phase}, this GL theory predicts a single-$Q$ to multiple-$Q$ phase boundary that is in good agreement with the full microscopic theory. In particular, substituting ${\Delta}({\bm R})=\Delta_qe^{iqx}+\Delta_{-q}e^{-iqx}$ into Eq.~\Eqref{func-i} (this is for a field applied along $\hat{y}$) and keeping powers to fourth order in the order parameter components yields the following expressions for the parameters in GL free energy
\begin{widetext}
\begin{align}\label{GLparam}
\kappa_{1x}&= \frac{\pi T}{8}v_F^2\sum_{n\ge 0}\Big\{N_+\frac{\omega_n^2-2H_q^2}{(\omega_n^2+H_q^2)^{5/2}}+N_-\frac{\omega_n^2-2H_{-q}^2}{(\omega_n^2+H_{-q}^2)^{5/2}}\Big\}
~,~
\kappa_{1y}= \frac{\pi T}{8}v_F^2\sum_{n\ge 0}\Big\{\frac{N_+}{(\omega_n^2+H_q^2)^{3/2}}+\frac{N_-}{(\omega_n^2+H_{-q}^2)^{5/2}}\Big\}
\nonumber \\
\kappa_{2x}&= \frac{\pi T}{8}v_F^2\sum_{n\ge 0}\Big\{N_-\frac{\omega_n^2-2H_q^2}{(\omega_n^2+H_q^2)^{5/2}}+N_+\frac{\omega_n^2-2H_{-q}^2}{(\omega_n^2+H_{-q}^2)^{5/2}}\Big\}
~,~
\kappa_{2y}= \frac{\pi T}{8}v_F^2\sum_{n\ge 0}\Big\{\frac{N_-}{(\omega_n^2+H_q^2)^{3/2}}+\frac{N_+}{(\omega_n^2+H_{-q}^2)^{5/2}}\Big\}
\nonumber \\
\alpha_1&= \frac{1}{V}-2\pi T\sum_{n\ge 0}\Big\{\frac{N_+}{\sqrt{\omega_n^2+H_q^2}}+\frac{N_-}{\sqrt{\omega_n^2+H_{-q}^2}}\Big\}
~,~
\beta_1= \frac{\pi T}{2}\sum_{n\ge 0}\Big\{N_+\frac{2\omega_n^2-H_q^2}{(\omega_n^2+H_q^2)^{5/2}}+N_-\frac{2\omega_n^2-H_{-q}^2}{(\omega_n^2+H_{-q}^2)^{5/2}}\Big\}
\nonumber \\
\alpha_2&= \frac{1}{V}-2\pi T\sum_{n\ge 0}\Big\{\frac{N_-}{\sqrt{\omega_n^2+H_q^2}}+\frac{N_+}{\sqrt{\omega_n^2+H_{-q}^2}}\Big\}
~,~
\beta_2= \frac{\pi T}{2}\sum_{n\ge 0}\Big\{N_-\frac{2\omega_n^2-H_q^2}{(\omega_n^2+H_q^2)^{5/2}}+N_+\frac{2\omega_n^2-H_{-q}^2}{(\omega_n^2+H_{-q}^2)^{5/2}}\Big\}
\nonumber \\
\beta_m&= 2\pi T\frac{(N_++N_-)}{qv_F}\sum_{n\ge 0}\Big\{\frac{(2\omega_n^2+H_q^2)H_q}{\omega_n^2(\omega_n^2+H_q^2)^{3/2}}  +\frac{(2\omega_n^2+H_{-q}^2)H_{-q}}{\omega_n^2(\omega_n^2+H_{-q}^2)^{3/2}}\Big\}
\end{align}
\end{widetext}
where $H_{q}=H-qv_F/2$, $H_{-q}=H+qv_F/2$, and $v_F$ is the Fermi velocity (note that this is equal for helicity bands). In the limit $\delta N=0$, this agrees with Ref.~\onlinecite{dim07}. We use this GL theory to examine the appearance of topological defects when an additional field is applied along the $c$-axis. Specifically, we take the in-plane Zeeman
field is fixed to be $H_y\mu_B/T_c=4.5$ and $\delta N=0.05$. For this in-plane field, $q=\mu_B H_y/mp_F>1/\xi_0$,  so that the use of the GL theory is well justified (since this theory describes spatial variations with length scales on the order of $\xi(T)\gg\xi_0$). We determine the parameters of the GL free energy for fixed in-plane field and by varying $0.1<T/T_c<0.156$ along the line shown in \Figref{Fig:phase}.

In \Figref{Fig:phase}, we also show the ground-state densities and the inverse of
the length scales $\xi_\pm$ and $\lambda$, respectively, on panel (B) and (C).
These are calculated from the Ginzburg-Landau functional \Eqref{GL-free} and
for the parameters obtained from the microscopic theory according to \Eqref{GLparam}.
The parameters of the Ginzburg-Landau theory \Eqref{GLparam} where calculated
for an in-plane field $H_y\mu_B/T_c=4.5$, in the temperature range $0.1<T/T_c<0.156$.
This corresponds to the region of the phase diagram labeled ``Investigated regime''
in panel (A) of \Figref{Fig:phase}.
Note that the coefficients $\kappa_{ax}$ and $\kappa_{ay}$ in front of the
kinetic terms in the Ginzburg-Landau model \Eqref{GL-free} are different.
For practical purposes, for example in numerical simulations, it is convenient
to absorb these anisotropies by a suitable rescaling of the fields and spatial
coordinates. These details are presented in Appendix~\Ref{Appendix}.

\section{Magnetic fields along \texorpdfstring{$c$}{c}-axis}
\label{Sec:C-axis}
%
Now we address the magnetic response of this system to applied field 
$H_z$ along the $c$-axis. To this end we minimize the microscopically 
derived free energy \Eqref{GL-free} for decreasing temperatures. That is, 
moving from right to left along the "Investigated regime" line shown in 
\Figref{Fig:phase}, panel (A). More precisely, the fields are discretized 
within a finite element formulation \cite{freefem}. Then, for a given value 
of the applied field $H_z$, the free energy is minimized at the temperature 
$T/T_c=0.16$. Once the minimization algorithm converges, the temperature 
is slightly decreased and the free energy is minimized again. This procedure 
is iterated until the temperature $T/T_c=0.10$ is reached.
Below, unless otherwise specified, the magnetic field $B$ is given in
units of $\Hc{2}^{(q)}$, and the condensate densities in units of their
ground state value. The spatial coordinates $x$ and $y$ are scaled in 
units $\sqrt{2\kappa_{1x}}$ and $\sqrt{2\kappa_{1y}}$ respectively.

\begin{figure}[!htb]
\hbox to \linewidth{ \hss
\includegraphics[width=0.8\linewidth]{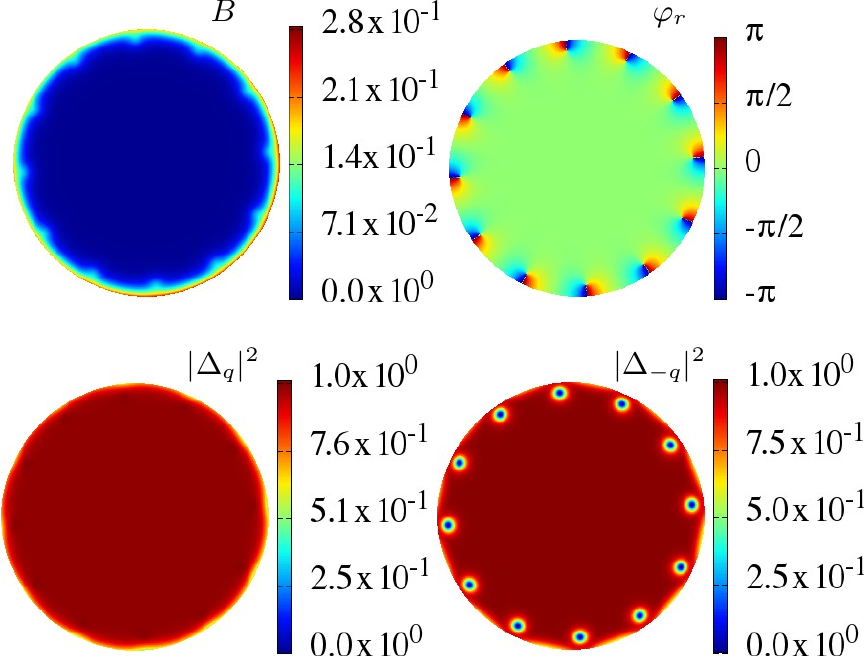}
\hss}
\caption{
(Color online) -- 
Stable fractional vortices near boundaries.
Displayed quantities are the magnetic field $B$ (in units of $\Hc{2}^{(q)}$),
the condensate densities in the units of their ground state value, and the 
relative phase between both condensates $\varphi_r:=\varphi_{-}-\varphi_{+}$.
The GL parameters are $e=0.001$, $T/T_c=0.110$ and the applied field
along the $c$-axis, $H_z$ corresponds to $120$ flux quanta going through 
the sample's area, if in the normal state.
The spatial coordinates $x$ and $y$ are scaled units of
$\sqrt{2\kappa_{1x}}$ and $\sqrt{2\kappa_{1y}}$ respectively.
Vortices enter in the $\Delta_{-q}$ condensate, while no vortices
enter in $\Delta_q$. This can be seen from the phase difference,
where winding in $\varphi_r$ indicates the existence of fractional
vortices.
}
\label{Fig:Fractional-vortex}
\end{figure}
\subsection{Fractional Vortices}
When the penetration depth $\lambda$ is the largest length scale 
($\lambda\gtrsim\xi_+,\xi_-$), depending on the applied field $H_z$, 
we find an unusual magnetic response featuring skyrmions.
The smoking gun of this unconventional state already manifests in low
applied fields, featuring the formation of thermodynamically stable
fractional vortices near the boundary (see \Figref{Fig:Fractional-vortex}).
\begin{figure}[!htb]
\hbox to \linewidth{ \hss
\includegraphics[width=0.8\linewidth]{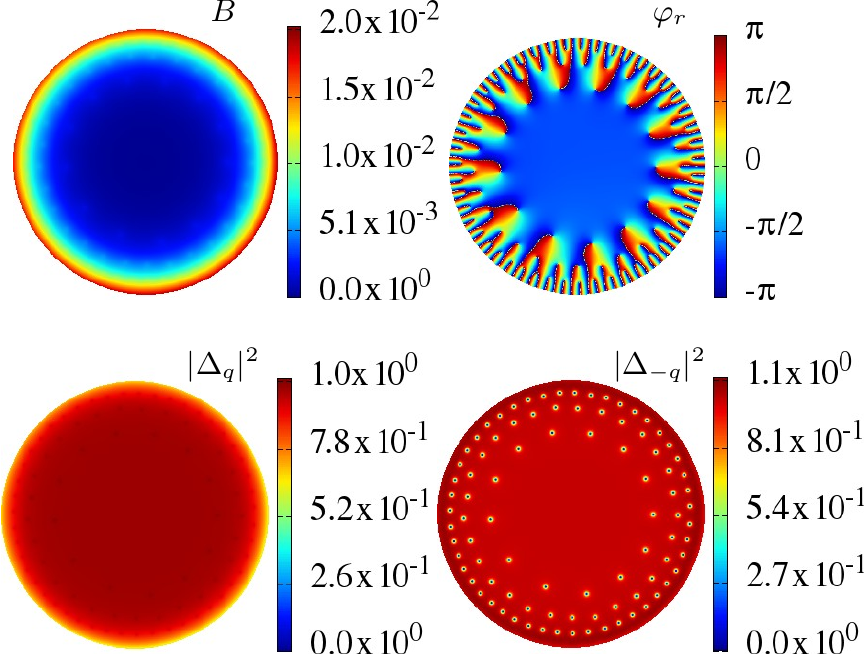}
\hss}
\caption{
(Color online)--  
Near boundary stable fractional vortices. Displayed quantities are
the same as in \Figref{Fig:Fractional-vortex}. The GL parameters are
$T/T_c=0.110$ and the applied field along the $c$-axis, $H_z$ corresponds
to $300$ flux quanta.
Quite remarkably, there are several layers of stable fractional
vortices near the boundary. Here the applied field is strong enough
to allow many (weak) vortices in $\Delta_{-q}$ to enter,
but not strong enough so that vortices in $\Delta_{q}$ can overcome the
Bean-Livingston barrier. If there are a few fractional vortices, they
would stay in a single layer closest to the boundary. However,
since there are so many fractional vortices, the system has
to compromise with multiple layers of fractional vortices.
}
\label{Fig:Fractional-vortex2}
\end{figure}
The mechanism for the stabilization of these boundary fractional vortices
\cite{sil11} can be understood as follows. The condition that no current
flows through the boundary is akin to placing an image fractional anti-vortex
outside the superconductor. The fractional vortex/anti-vortex pair does
not have a logarithmically divergent energy. Moreover, fractional vortices
in $\Delta_{-q}$ experience a smaller Bean Livingston barrier than
vortices in $\Delta_q$. Consequently, these fractional vortices have
the lowest field for entry into the superconductor. However, because they 
have logarithmically divergent energy in the bulk, they cannot deeply 
penetrate into the superconductor. To compromise between the 
vortex/(image)anti-vortex attraction and repulsion due to Meissner current, 
the fractional vortices sit at a preferred finite distance from the boundary 
\cite{sil11}, as can be seen in \Figref{Fig:Fractional-vortex}. It is rather 
interesting to note that several layers of fractional vortices can form as 
shown in \Figref{Fig:Fractional-vortex2}. The cost of fractional vortices 
increases with the depth of the layer. However, a multi-layered structure
of fractional vortices does form as long as the external field is weak enough 
(this is consistent with the results of Ref.~\onlinecite{sil11}).

\begin{figure*}[!htb]
\hbox to \linewidth{ \hss
\includegraphics[width=0.8\linewidth]{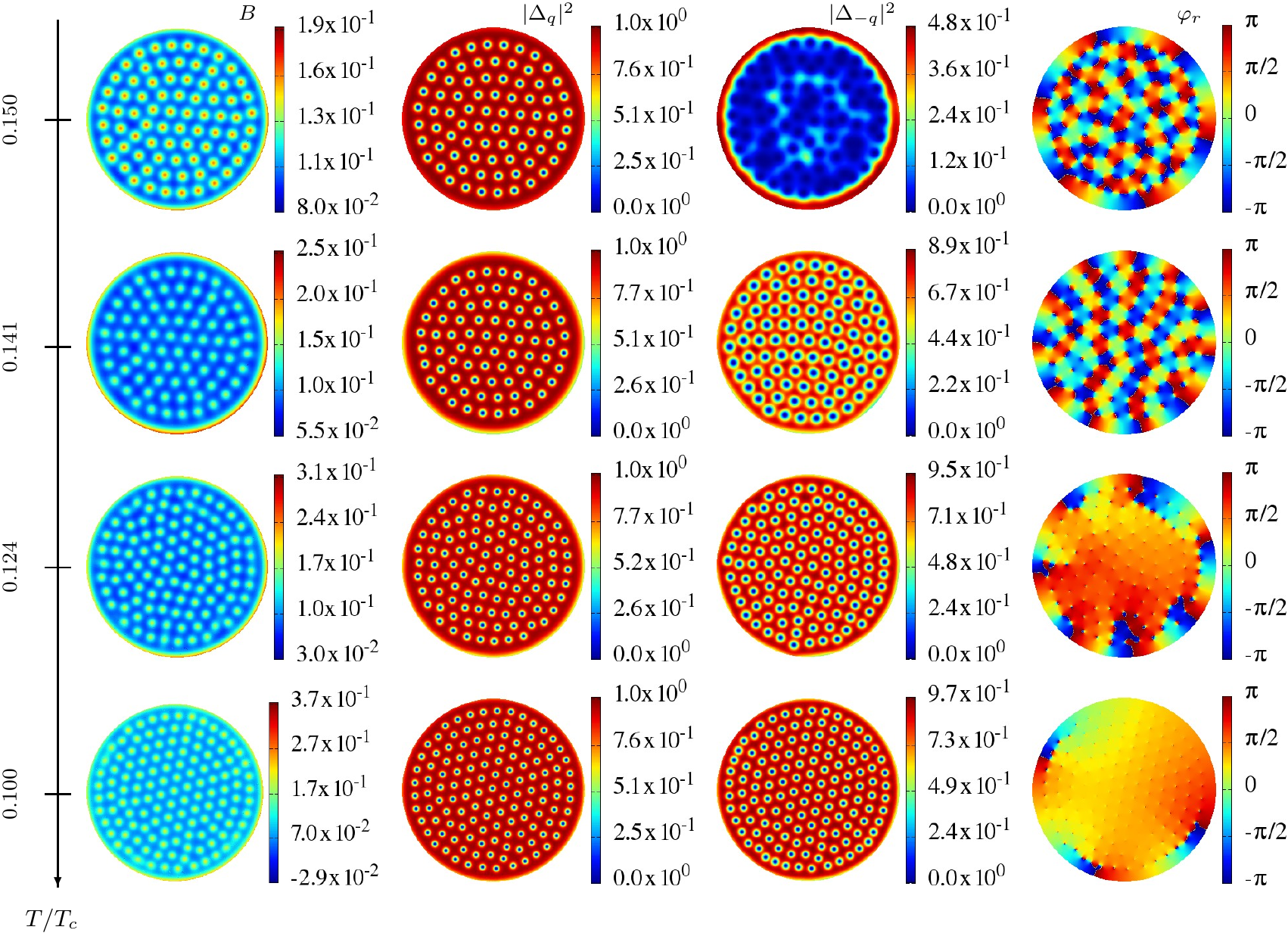}
\hss}
\vspace{-0.4cm}
\caption{
(Color online) -- 
Transition from a skyrmion lattice state to a vortex lattice state as 
temperature is decreased. This shows the magnetic field $B$ (in units of 
$\Hc{2}$), the condensate densities in the units of their ground state 
value, and the relative phase $\varphi_r=\varphi_{-}-\varphi_{+}$ between 
both condensates $\Delta_{q}$ and $\Delta_{-q}$.
The spatial coordinates $x$ and $y$ are scaled units of
$\sqrt{2\kappa_{1x}}$ and $\sqrt{2\kappa_{1y}}$ respectively.
Here, the applied field  exceeds the second critical field associated
with the (weak) component $\Delta_{-q}$. As a result, at $T/T_c=0.16$,
$\Delta_{-q}$ is completely suppressed. On the other hand, the field is 
strong enough for the component $\Delta_{q}$ to develop a regular triangular 
vortex lattice (here it is not perfect because the circular boundary causes
disclinations in the lattice).
As the temperature is decreased, $\Delta_{-q}$ develops superconductivity 
accompanied with many vortices. In this regime, close to the second critical 
field associated with the (weak) component $\Delta_{-q}$, the lattice in 
$\Delta_{-q}$ does not coincide with that in $\Delta_{q}$. This can be seen 
by the relative phase $\varphi_r$ of the two condensates.
With cooling further, as $\Delta_{-q}$ becomes larger, the skyrmion lattice 
is no longer favoured and vortices is both components tend to overlap. At the 
lowest temperatures, all vortices are co-centred and form a triangular lattice, 
apart from three extra fractional vortices in $\Delta_{-q}$ that are in a 
(meta-)stable near boundary configuration.
}
\label{Fig:Skyrmion-lattice-melting}
\end{figure*}
\subsection{Skyrmion lattice}

For higher values of the external field, vortices in $\Delta_q$ also overcome 
the surface Bean-Livingston barrier and start penetrating into the bulk. This 
results in formation of skyrmions, that are bound state of fractional vortices 
in both condensates. Since they carry integer flux, they are not confined to 
localize near the boundary. Yet, they can coexist with near-boundary fractional 
vortices. At high fields, as shown in \Figref{Fig:Skyrmion-lattice-melting}, 
lattices of vortices with split cores (as revealed by local winding in $\varphi_r$) 
are formed.
The density-density interaction term, controlled by $\beta_m$, is responsible
for the core-splitting of vortices. These split-core vortices, the skyrmions, 
carry integer flux and have finite energy. Moreover, because their cores are 
split, skyrmions are characterized by an additional invariant due to the 
non trivial homotopy $\pi_2(S^2)\in\mathbb{Z}$ of the maps $S^2\to S^2$. 
This can be understood by introducing a pseudo-spin unit vector ${\bf n}$ that
is the projection of the superconducting condensates onto spin-$1/2$ Pauli
matrices $\bs\sigma$:
\Equation{Projection}{
 {\bf n}=\frac{\Psi^\dagger\bs \sigma\Psi}{\Psi^\dagger\Psi}\,,
 ~~\text{where}~~
 \Psi^\dagger=(\sqrt{\kappa_{1x}}\Delta_q^*,\sqrt{\kappa_{2x}}\Delta_{-q}^*)\, .
}
This unit vector spans a two-sphere. Thus ${\bf n}$ is a map from
$S^2$ (the compactification of the real plane) to $S^2$. Such maps
are characterized homotopy class $\pi_2(S^2)\in\mathbb{Z}$. For a more
detailed discussion, see Appendix~\ref{Appendix}. Importantly, this
additional homotopy invariant is zero for Abrikosov vortices. Non trivial
texture of the pseudo-spin unequivocally indicates non-trivial topology
of the $S^2\to S^2$ map. This motivates the terminology \emph{skyrmion}.

Near the critical field $\Hc{2}^{(-q)}$, that is, just below the single-$Q$
to multiple-$Q$ transition, two condensates form triangular lattices that
are displaced with respect to each other. The displacement is such that vortices
of $\Delta_{-q}$ lie in the center of a triangle of vortices in $\Delta_q$.
While the temperature is decreased, as in \Figref{Fig:Skyrmion-lattice-melting},
vortices in $\Delta_{-q}$ no longer sit in the center of triangles of vortices
in $\Delta_{q}$. Vortices in $\Delta_{-q}$ pair with one of the three
surrounding vortices of $\Delta_{q}$ and the resulting skyrmion lattice
is no longer triangular. Further reducing the temperature induces stronger
binding of the pairs that eventually merge to Abrikosov vortices, at the
same time losing non-trivial features of the $S^2\to S^2$ map. The resulting
vortices form a usual triangular lattice. Similar behavior of the merging of
the skyrmion lattice into Abrikosov lattice occurs at fixed temperature with
decreasing fields.
\begin{figure}[!htb]
\hbox to \linewidth{ \hss
\includegraphics[width=0.8\linewidth]{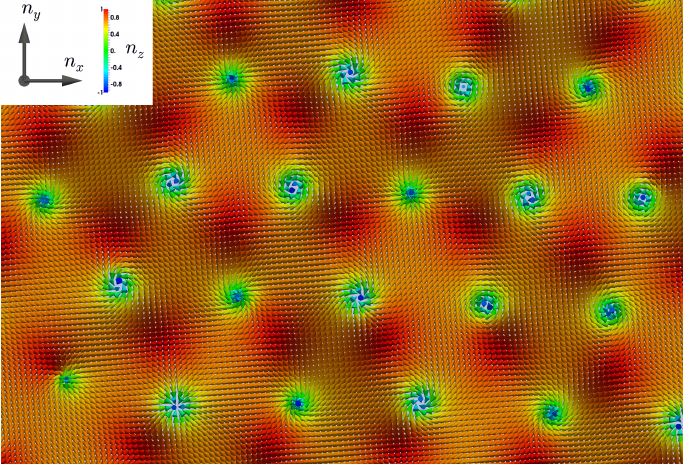}
\hss}
\caption{
(Color online) -- 
A skyrmion lattice.
This shows pseudo-spin (texture) ${\bf n}$ \Eqref{Projection}
obtained by projecting the superconducting condensates on Pauli
matrices.
It corresponds to the configuration shown on the second line of
\Figref{Fig:Skyrmion-lattice-melting}.
}
\label{Fig:Skyrmion-lattice-nfield}
\end{figure}
\Figref{Fig:Skyrmion-lattice-nfield} shows the typical texture field 
${\bf n}$ associated with a skyrmion lattice.
The core splitting strongly depends on the penetration depth. The larger
$\lambda$ (small $e$) is, the weaker the binding of fractional vortices becomes.
Skyrmion lattices should thus be easily identifiable in extreme type-2 
superconductors.


\subsection{Type-1.5 Superconductivity}
\label{type15}

For skyrmions to form, the penetration depth should be the largest
length scale. If this is not the case, vortices will not split their 
cores and thus co-centered vortices, which are trivial regarding
$S^2\to S^2$ maps, are formed.
In the regime where the penetration depth $\lambda$ is not the largest 
length scale, in addition to thermodynamically stable fractional vortices 
near boundaries,  different unconventional features can arise. These 
features are associated with the fact that the coherence lengths
$\xi_-$ diverges at the transition between single-$Q$ and multiple-$Q$
phase, while both the coherence length $\xi_+$ and the penetration depth
$\lambda$ remain finite (see panel (C) of \Figref{Fig:phase}).
Moreover if $\xi_+<\lambda$, then there always exist a temperature range
where $\xi_+<\lambda<\xi_-$. Such a regime is termed ``type-1.5'' \cite{moshchalkov} 
and is a subject of great experimental interest \cite{ray}. In this regime, 
vortices are favoured over skyrmions. Interactions between vortices are 
non-monotonic long-range attractive and short-range repulsive
\cite{Babaev.Speight:05,Carlstrom.Babaev.ea:10,silaev2011}.
\begin{figure}[!htb]
\hbox to \linewidth{ \hss
\includegraphics[width=\linewidth]{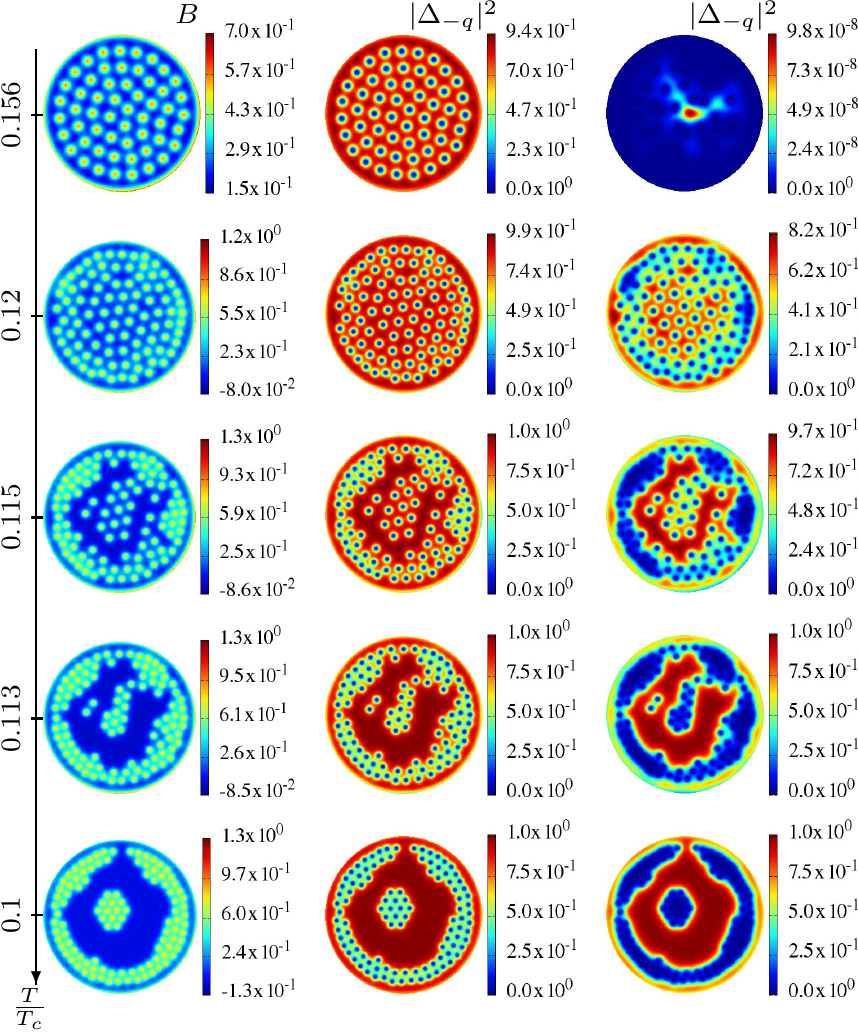}
\hss}
\vspace{-0.2cm}
\caption{
(Color online) -- 
Vortex aggregation at fixed field with cooling temperature.
This shows the magnetic field $B$ and the condensate densities.
At $T/T_c=0.156$, $\Delta_{-q}$ is completely suppressed, while
$\Delta_{q}$ develops a triangular lattice up to some disclinations
due  the domain geometry (the disk cannot perfectly accommodate the
lattice).
As the temperature is decreased, $\Delta_{-q}$ develops
superconductivity accompanied by vortices.
When further cooled, vortices coalesce into clusters. The attraction
between vortices is mediated by the (hybridized) coherence length
$\xi_-$, which has longest range. The long-range attraction gets
stronger as the system is cooled and vortex aggregates become very
compact.
}
\label{Fig:Aggregation}
\end{figure}
Due to the preferred intervortex separation, the usual Abrikosov lattice
coalesces into vortex aggregates as shown in \Figref{Fig:Aggregation}.
At temperatures close to $T/T_c=0.16$ (the single-$Q$ to multiple-$Q$
transition temperature when $H_z=0$), the external field exceeds the second
critical field $\Hc{2}^{(-q)}$ of the second condensate and $\Delta_{-q}$
is completely suppressed. A moderate applied field nevertheless does not
destroy $\Delta_{q}$ and it shows an hexagonal lattice of vortices. While
the system is cooled, $\Delta_{-q}$ develops superconductivity accompanied
with the formation of vortices. The disparity in length scales, leads to
long-range attractive force between vortices.
Initially, the attractive tail is rather weak and the effect on vortices is
hardly noticeable. With further cooling, as $\Delta_{-q}$ increases,
the attractive interaction becomes much stronger, seeding inhomogeneities
in the vortex distribution to finally lead to segregation of clusters
of vortices surrounded by regions of Meissner state.
It is worth emphasizing here that clusters tend to sit near the boundary
rather than in the bulk.

\section{Conclusions}
Our findings reveal that clean $s$-wave interface superconductors with 
strong Rashba spin-orbit coupling should exhibit novel and interesting 
physics under magnetic fields. The predicted skyrmion and fractional vortex 
formation and  can be observable through scanning SQUID and scanning 
tunnelling microscopy measurements.
We also have shown that the system should fall into the regime of type-1.5 
superconductivity for some parameter range, however a reservation should 
be made that due to large magnetic field penetration length in these materials, 
this regime may only exist for a small window of temperatures. The observation 
of these effects will open a new window into the relationship between topological 
defects and magnetic response in superconductors.

\begin{acknowledgments}
We thank the Aspen Center for Physics where this work was initiated (NSF Grant No. 1066293). DFA acknowledges support from NSF grants No. DMR-0906655 and No. DMR-1335215. EB acknowledges support  by the Knut and Alice Wallenberg Foundation through a Royal Swedish Academy of Sciences Fellowship, by the Swedish Research Council, and by the National Science Foundation under the CAREER Award DMR-0955902.
The computations were performed on resources provided by the Swedish National Infrastructure for Computing (SNIC) at National Supercomputer Center at Link\"oping, Sweden.
\end{acknowledgments}

\appendix
\section{Properties of the Ginzburg-Landau model}
\label{Appendix}

Because the coefficients in front of the kinetic term are different,
$\kappa_{ax}\neq\kappa_{ay}$, the Ginzburg-Landau model \Eqref{GL-free}
is anisotropic. To analyze the properties of the Ginzburg-Landau model,
it is convenient to absorb the anisotropies by rescaling the fields
and spatial coordinates. First, note that we take the prefactors of
the kinetic terms to satisfy the relation
$\kappa_{1y}/\kappa_{1x}=\kappa_{2y}/\kappa_{2x}$ which is numerically
found to be accurate to $10^{-4}$. The anisotropies are absorbed using
the following parametrization
\Equation{AppAnisotropies}{
\left\lbrace \begin{array}{c}
\kappa_{1x}=\kappa_1/2 \\
\kappa_{1y}= m^2\kappa_1/2
\end{array}\right.
~~~~~
\left\lbrace \begin{array}{c}
\kappa_{2x}=\delta\kappa_{1x} \\
\kappa_{2y}=\delta\kappa_{1y}
\end{array}\right.\,.
}
The rescaled (isotropic) spatial coordinates are
\Equation{AppCoordinates}{
\tilde{x}=x/\sqrt{\kappa_1} \,,~~~ \tilde{y}=y/m\sqrt{\kappa_1} \,,
}
and the rescaled fields
\Align{AppGF}{
\tilde{A}&=A_x/m\sqrt{\kappa_1} \,,~~~
\tilde{A}_y=A_y/\sqrt{\kappa_1} \,,~~~ \nonumber \\
\psi_1&=\Delta_q \,,~~~
\psi_2=\sqrt{\delta}\Delta_{-q}
\,.
}
Defining the new parameters of the interacting potential
\Align{AppParameters}{
\tilde{\alpha}_1&=\alpha_1\,,~~
\tilde{\beta}_1=\beta_1\,,~~	
\tilde{\alpha}_2=\frac{\alpha_2}{\delta}\,,~~
\tilde{\beta}_2=\frac{\beta_2}{\delta^2}\,,~~ \nonumber \\
\tilde{\gamma}&=\frac{\beta_m}{\delta}\,,~~
\tilde{e}=2em\kappa_1\,,
}
and defining the rescaled free energy $\tilde{\mathcal{F}}=m\kappa_1f$,
the Ginzburg-Landau model \Eqref{GL-free} now reads (now on we omit the
$\tilde{\,}$ symbols)
\Align{AppFreeEnergy}{
 \mathcal{F}&= \sum_{a}\Big\{\frac{1}{2}|(\Grad+ie\A)\psi_a|^2   	
+\alpha_a|\psi_a|^2+\frac{1}{2}\beta_a|\psi_a|^4 	\Big\} \nonumber\\
+&\gamma|\psi_1|^2|\psi_2|^2+\frac{1}{2}\B^2-B_zH_z
 \,,
}
and the magnetic field is $\B=\Curl\A$.
Functional variation of the free energy functional \Eqref{AppFreeEnergy}
determines the Euler-Lagrange equations of motion. That is, variation
with respect to complex fields $\psi_a^*$ gives the Ginzburg-Landau
equation for the condensates, while variation with respect to the
vector potential defines the Amp\`ere's law
\Align{AppEOM}{
	\D\D\psi_a&=2\frac{\partial V(\Psi)}{\partial\psi_a^*}\,, ~~~
	\Curl\B+\J=0	\,,\nonumber \\
\text{and}~~~&
  \J\equiv\sum_{a}\J^{(a)}=
   \sum_{a}e\Im\left(\psia^*\D\psia  \right)\,,
}
where the covariant derivative is $\D\equiv\Grad+ie\A$.
The ground state is the state with constant densities of the
superconducting condensates $|\psi_a|=u_a$, while $\A$ is a pure gauge
that can be chosen to be zero. Ground state densities satisfy
\Equation{AppGS}{
\left\{ \begin{array}{c}
   \left(\alpha_1+\beta_1u_1^2+\gamma u_2^2\right)u_1 =0\\
   \left(\alpha_2+\beta_2u_2^2+\gamma u_1^2\right)u_2 =0
\end{array}\right.
}
They are defined up to an arbitrary global phase, which can be set to 
zero without loss of generality since the free energy \Eqref{AppFreeEnergy} 
is invariant under global \groupUU transformation.
Provided $\beta_1\beta_2-\gamma^2>0$ \emph{and} $\alpha_1,\alpha_2<0$, 
if $\alpha_2\gamma-\alpha_1\beta_2>0$ and $\alpha_1\gamma-\alpha_2\beta_1>0$,
both ground state densities $u_1$, $u_2$ are non zero:
\Equation{AppGSsol}{
 u_1^2=\frac{\alpha_2\gamma-\alpha_1\beta_2}{\beta_1\beta_2-\gamma^2}\,, ~~~
 u_2^2=\frac{\alpha_1\gamma-\alpha_2\beta_1}{\beta_1\beta_2-\gamma^2}\,.
}
The conditions $\beta_1\beta_2-\gamma^2>0$, $\alpha_1,\alpha_2<0$ are 
necessary conditions for both condensates to have non zero ground states 
and $\alpha_2\gamma-\alpha_1\beta_2>0$, $\alpha_1\gamma-\alpha_2\beta_1>0$
are the stability conditions of the ground state. Both are verified for
the range of temperature we considered.

The length scales at which a perturbed condensate recovers its
ground-state density, the coherence lengths $\xi_a$, are determined
by expanding the fields around the ground state $\psi_a=u_a+\epsilon_a$
and linearising the equations of motion \Eqref{AppEOM}. The length scales
are determined by the eigenvalues $\mathcal{M}_a^2$ of the mass matrix
\Equation{AppMassMatrix}{
\mathcal{M}^2=\left( \begin{array}{c c}
\alpha_1+3\beta_1 u_1^2+\gamma u_2^2		&2\gamma u_1u_2 \\
2\gamma u_1u_2							&\alpha_2+3\beta_2 u_2^2+\gamma u_1^2
\end{array}\right)\,,
}
and the coherence lengths are $\xi_a=1/\sqrt{2\mathcal{M}_a^2}$, while the
penetration depth is $e\lambda=1/\sqrt{u_1^2+u_2^2}$. The ground-state
densities and length scales, for the values of the Ginzburg-Landau
potential parameters derived from the microscopic calculations, are
displayed in \Figref{Fig:phase}.

Note that the coherence lengths $\xi_1$ and $\xi_2$ are not the coherence
lengths associated with each condensates independently. Indeed, the modes
here are hybridized. That is, $\xi_1$ and $\xi_2$ are the coherence lengths
associated to orthogonal linear combinations of the fields $\psi_1$ and
$\psi_2$. These linear combinations form the eigenbasis of the mass matrix
\Eqref{AppMassMatrix}.

\subsection{Flux quantization and fractional vortices}

The elementary topological excitations, when several condensates
are involved, are fractional vortices. These are field configurations
with $2\pi$ phase winding of only one condensate. For example, $\varphi_1$
has $\oint\Grad\varphi_1=2\pi$ winding while $\oint\Grad\varphi_2=0$.
A fractional vortex in the condensate $a$, carries only a fraction of
flux quantum. This can be seen by calculating the magnetic flux.
The supercurrent, defined from the equations of motion \Eqref{AppEOM},
reads as
\Equation{AppCurrents}{
\J/e:=\frac{\delta\mathcal{F}}{\delta\A}=
  e\varrho^2\A+\sum_{a}|\psi_a|^2\bs\nabla\varphi_a \,.
}
Here we defined the total density  $\varrho^2=\sum_a|\psi_a|^2$.
Since the supercurrent $\J$ is screened, it decays exponentially
and there, the condensates have constant density. The magnetic flux
thus reads as
\Align{AppFlux}{
\Phi&=\int\B\cdot\bs{dS}=\oint \A \cdot\bs{d\ell} 	\nonumber\\
	&=\frac{1}{e^2\varrho^2}
	\oint \left(\J-e\sum_{a}|\psi_a|^2\Grad\varphi_a\right)
	\cdot\bs{d\ell} \nonumber\\
	&=-\frac{\sum_a|\psi_a|^2}{e\varrho^2}
	\oint\Grad\varphi_a \cdot\bs{d\ell}\,.
}
Each condensate must wind an integer number of times. Thus,
if $\psi_1$ winds $n_1$ times and $\psi_2$ winds $n_2$ times,
the flux reads as \cite{bab02}
\Equation{AppFlu2x}{
\Phi=n_1\frac{|\psi_1|^2\Phi_0}{|\psi_1|^2+|\psi_2|^2}
	+n_2\frac{|\psi_2|^2\Phi_0}{|\psi_1|^2+|\psi_2|^2}
	\,,
}
where the flux quantum is $\Phi_0=2\pi/e$. Fractional vortices
in different condensates attract each other logarithmically at long
distances. If $n_1=n_2$, the two fractions
$\frac{|\psi_a|^2}{|\psi_1|^2+|\psi_2|^2}$ of flux add up to give integer
flux. Restoring the original unit system gives the flux in Eq.~\Eqref{flux} of
the main text.

\subsection{Topology}

Vortices, either fractional or composite are characterized by $S^1\to S^1$
topological maps. The first circle $S^1$ denotes the closed path faraway
from the vortex core (that is homeomorphic to a circle) while the second
one (the target circle) correspond to \Uone rotations. Heuristically the
$S^1\to S^1$ maps have the following meaning: they count how many times the
target circle is covered while going along the closed path faraway
from the vortex core. That is the number of phase windings. Importantly,
this number can be calculating just by inspecting the closed path faraway
for the vortex core. This is because the associated density of the topological
invariant is a total divergence.

There are two kinds of field configurations that carry an integer number of
flux quanta. Ordinary vortices, for which the two components wind around the
same point, and skyrmions for which the two components do not wind around the 
same point. In the model that we derived, these can continuously be deformed 
into each other at a finite free energy cost. And one kind of topological defect 
is unstable against decaying into another.
The vortices and skyrmions can be distinguished on topological grounds,
by the topological invariant associated with $S^2\to S^2$ topological maps.
The first $S^2$ here stands for the compactification of the $\mathbb{R}^2$
plane. Heuristically this topological map counts the number of times the
target sphere (defined below as the projection of the condensates on Pauli
matrices), is covered while covering the plane $\mathbb{R}^2$.
Note that here the density of topological invariant associated with this
map is not a total divergence (see Eq.~\Eqref{AppCharge}). Thus the topological
invariant, which cannot be reduced to a line integral, is clearly different
from that of the $S^1\to S^1$ maps.

The topological invariant is rigorously derived by defining $\Psi$,
the vector of complex fields: $\Psi^\dagger=(\psi_1^*,\psi_2^*)$.
$\Psi$ is a smooth holomorphic map $M\rightarrow\mathbb{C}^2$ with the
manifold $M$ denoting the infinite plane $\mathbb{R}^2$. Note that $M$ can
also stand for the one-point compactification of the plane, which is homeomorphic
to a sphere $M=\mathbb{R}^2\cup\{\infty\}\simeq S^2$. Now, we define the
projection $\pi:\mathbb{C}^2\backslash\{0\}\rightarrow\groupCP{1}$, which is
roughly speaking the stereographic projection. Then $\phi=\pi\circ\Psi$
is a $\groupCP{1}$-valued field that maps all points in $M$ to a point in
$\groupCP{1}$. This has topological degree
$\Q(\Phi)=\frac{1}{4\pi}\int_M\phi^*\omega\in\mathbb{Z}$, where $\omega$ is
the K\"ahler form associated with the Fubini-Study metric on $\groupCP{1}$.
The topological index $\Q$ can be rewritten in terms of the complex field
$\Psi$ as
\Equation{AppCharge}{
  \Q(\Psi)=\int_{M}\frac{i\epsilon_{ji}}{2\pi|\Psi|^4} \left[
   |\Psi|^2\partial_i\Psi^\dagger\partial_j\Psi
   +\Psi^\dagger\partial_i\Psi\partial_j\Psi^\dagger\Psi
   \right]d^2x\,,
}
and $\Q$ equals the number of enclosed flux quanta ($2\pi\Q/e=\int_MBdS$),
provided $\Psi\neq0$. For a more complete, rigorous and detailed demonstration,
see the demonstration for an arbitrary number of complex fields in
Ref.~\onlinecite{Garaud.Carlstrom.ea:13}.
The \CPone topological invariant \Eqref{AppCharge} for skyrmions can alternatively
be derived using the pseudo-spin ${\bf n}$. That derivation is equivalent to the 
derivation we used for \Eqref{AppCharge}. The pseudo-spin unit vector $\bf n$ 
is the projection of superconducting condensates on spin-$1/2$ Pauli matrices 
$\bs\sigma$:
\Equation{AppProjection}{
 {\bf n}\equiv (n_x,n_y,n_z)
 =\frac{\Psi^\dagger\bs \sigma\Psi}{\Psi^\dagger\Psi}\,,
 ~~\text{where}~~
 \Psi^\dagger=(\psi_1^*,\psi_2^*)\, .
}
Roughly speaking this is the stereographic projection $\pi$ mentioned earlier.
The pseudo-spin is a map from the one-point compactification of the plane
($\mathbb{R}^2\simeq S^2 $) to the two-sphere target space spanned by $\bf n$.
That is ${\bf n}: S^2\to S^2$, and this map is characterized by the homotopy
class $\pi_2(S^2)\in\mathbb{Z}$, thus defining the integer valued topological
(skyrmionic) charge
\Equation{AppChargeBis}{
   \Q({\bf n})=\frac{1}{4\pi} \int_{\mathbb{R}^2}
   {\bf n}\cdot\partial_x {\bf n}\times \partial_y {\bf n}\,\,
  dxdy \,.
}
The key point in both these derivations \Eqref{AppCharge} and \Eqref{AppChargeBis}
of the integer topological invariant $\Q$, is that it relies on the fact
that $\Psi\neq0$. That is it is integer only if $\psi_1$ and $\psi_2$ have
no coincident cores. In other words, ordinary (composite) vortices with a
single core $\Psi=0$, have $\Q=0$. Skyrmions (core-split vortices), on the
other hand, have non-trivial charge $\Q=N$ (with $N$ coincides with the number
of carried flux quanta).
Note that when $\Psi\neq0$, $\Q$ is an integer provided the fields recover
ground state at the boundary of the integration domain. This means that,
for example, Meissner currents make this definition for $\Q$ non integer.

%

\end{document}